\documentclass[twocolumn,amsmath,amssymb,superscriptaddress]{revtex4-2}
\usepackage[utf8]{inputenc}
\usepackage{natbib}
\usepackage{graphicx}
\usepackage{amsmath}
\usepackage{amssymb}
\usepackage{amsthm}
\usepackage{comment}
\usepackage[usenames,dvipsnames]{xcolor}

\begin{document}

\title{Capillary self-folding chains}

\author{Megan~Delens}
\email{Corresponding author: megan.delens@uliege.be}
\affiliation{GRASP, Physics Department, Universit\'e de Li\`ege, Belgium.}
\author{Axel~Franckart}
\affiliation{GRASP, Physics Department, Universit\'e de Li\`ege, Belgium.}
\author{Martin~Poty}
\affiliation{GRASP, Physics Department, Universit\'e de Li\`ege, Belgium.}
\author{Nicolas~Vandewalle}
\affiliation{GRASP, Physics Department, Universit\'e de Li\`ege, Belgium.}


\begin{abstract}
Mesoscale self-assembly provides a route toward the design of
programmable microsystems. Here, we construct flexible chains of floating monomers whose curved branches impose upward or downward deformations of the liquid interface, corresponding to effective positive or negative capillary charges. These geometrically encoded deformations generate local attractive or repulsive interactions along the chain. By tuning the capillary sequence, we obtain distinct folded configurations, including straight lines, zigzag patterns, and loops. For short chains, folding is largely governed by nearest-neighbor
interactions and leads to well-defined structures. As the chain length increases and non-neighboring segments come into proximity and interact, however, the folding landscape becomes increasingly complex, with multiple metastable states whose number grows exponentially with chain length. We map these landscapes numerically and demonstrate experimentally that mechanical agitation allows the chains to transition between metastable configurations. Beyond encoding
a target geometry, the capillary sequence therefore controls the complexity of the folding landscape as well as the degeneracy and mutational robustness of folded structures. These results establish capillary chains as a controllable mesoscale platform for investigating how local interaction rules give rise to collective folding and complex sequence-to-structure relationships reminiscent of those encountered in biomolecular systems.
\end{abstract}

\maketitle

\section*{Introduction}
The spontaneous folding of a linear chain into a well-defined structure, driven solely by its sequence, is one of nature's most elegant design principles. In proteins, amino acid sequences encode three-dimensional conformations through hydrophobic, electrostatic, and hydrogen-bonding interactions~\cite{anfinsen1973principles,bryngelson1995funnels,leopold1992protein}. 
In DNA origami, complementary base pairing directs the formation of programmable 2D and 3D nanostructures~\cite{seeman1982nucleic,winfree1998design,rothemund2006folding}. 
In both cases, a discrete alphabet encodes interaction rules that strongly constrain the resulting structure.
Sequence-programmed folding has also been explored at mesoscopic and macroscopic scales. Millimeter-scale charged beads threaded on a flexible string provide a physical beads-on-a-string model of sequence-dependent folding \cite{reches2009beads}. DNA-functionalized emulsion droplets have subsequently enabled controlled bond valence, sequential assembly, and the formation of freely jointed colloidal polymers~\cite{feng2013specificity,zhang2017sequential,mcmullen2018freely}.
Most directly, McMullen et al. demonstrated that the sequence of a droplet chain and the order in which its secondary interactions are activated can be designed to produce specific two-dimensional foldamers~\cite{mcmullen2022programmable}.
In parallel, theoretical studies have established how local interaction
rules, alphabet size, competing excited states, and interaction crosstalk constrain the yield and information capacity of programmable self-assembly~\cite{hormoz2011design,zeravcic2014size,murugan2015multifarious,huntley2016capacity}.
These developments form part of a broader effort to engineer colloidal matter with programmable and life-like behavior~\cite{zeravcic2017living}.
This raises a complementary question: how can a minimal, static, geometry-based interaction alphabet control not only a target conformation, but the metastable folding landscape of a connected chain? Capillary forces at a liquid interface provide a natural route to this problem. Floating particles deform the interface according to their geometry and wetting properties, generating menisci that attract or repel neighboring objects~\cite{nicolson1949,vella2005,ho2019direct,hooshanginejad2024interactions}. The sign and magnitude of these interactions are captured by the concept of a capillary charge~\cite{chan1981,kralchevsky2001capillary}. Particles that generate menisci of the same sign attract each other, whereas opposite menisci repel. In this sense, capillary charges play a role analogous to that of electric charges, while remaining entirely governed by interfacial deformation. With anisotropic particles, the contact line undulates along the particle shape, creating multipolar effects and directional interactions~\cite{botto2012capillaryreview,hosokawa1996two,danov2010capillary}. 
This geometrical encoding has been exploited to assemble ordered structures~\cite{bowden1997self,loudet2005capillary,poty2014customizing,
delens2025encoding}, enable reconfigurable assemblies~\cite{davies2015dipolar,vandewalle2020switchable,metzmacher2017self}, 
and achieve programmable manipulation~\cite{delens_3d-printed_2025,metzmacherring_2025,LIU2022132380}, establishing capillarity as a versatile route toward programmable matter \cite{whitesides2002self}.

\begin{figure}[ht]
\includegraphics[width=\linewidth]{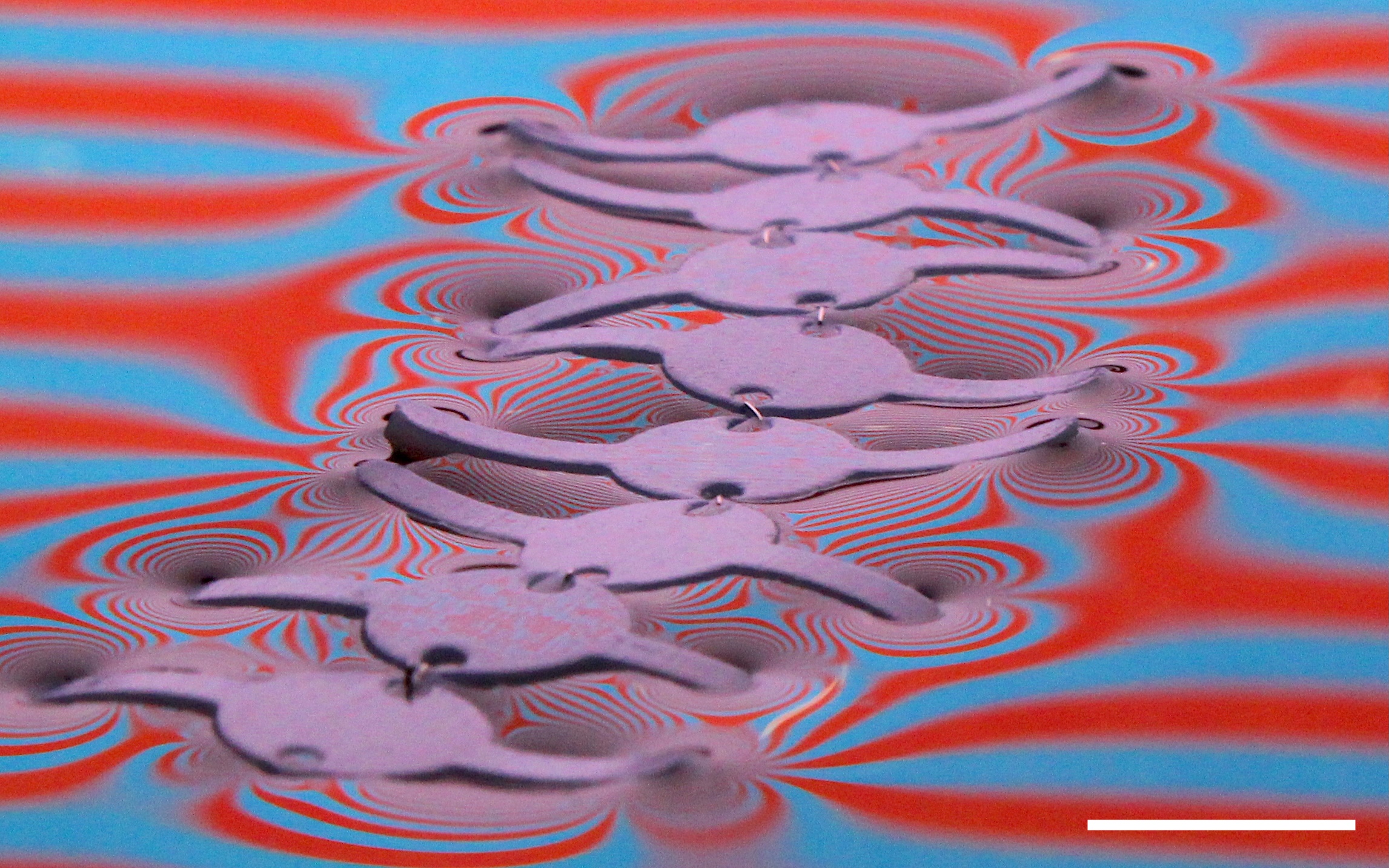}
\caption{Picture of a flexible chain made of 8 components floating on a liquid-air interface. Each component has two opposite branches. By using the reflection of a striped pattern, the local deformations of the liquid interface are evidenced. Negative and positive curvatures are visible near branch tips. Scale bar: 10 mm.}
\label{fig:chain}
\end{figure}

Here, we introduce a mesoscale platform in which capillary interactions encode the folding of flexible chains floating at a liquid interface. The system, illustrated in Figure~\ref{fig:chain}, consists of elementary floating units whose geometry encodes positive and negative capillary charges, thereby defining local attractive and repulsive interactions along the chain. In this system, the interaction rules are fixed by the monomer geometry and are simultaneously present throughout the folding process. Combining experiments and numerical modeling, we first establish how these local capillary codes determine the stable states of individual joints, and then investigate how they collectively shape the folding landscape of longer chains. As the chain grows, segments that are distant along the backbone can come into close proximity and interact, generating multiple competing metastable configurations. We quantify how the number and organization of these states evolve with chain length, experimentally drive transitions between them using mechanical agitation, and finally examine how the capillary sequence controls structural degeneracy and mutational robustness. Capillary chains thus provide a controllable mesoscale platform for investigating how simple, geometrically encoded interaction rules give rise to collective folding and complex sequence-to-structure relationships.

\section*{Results and discussion}
\subsection*{Capillary code in one monomer}
The self-folding behavior of the chain originates from the capillary information encoded in each of its monomers. Previous studies have shown that floating objects with pinned contact lines can encode capillary interactions through their geometry: the curvature of their edges generates multipolar interface deformations whose symmetry and strength can be precisely programmed~\cite{brown2000fabricating,poty2014customizing,bae2017programmable,metzmacher2017self,vandewalle2020switchable,delens2025encoding}. Building on this principle, we design a capillary monomer carrying two independently programmable curved branches.

\begin{figure}[ht]
\includegraphics[width=\linewidth]{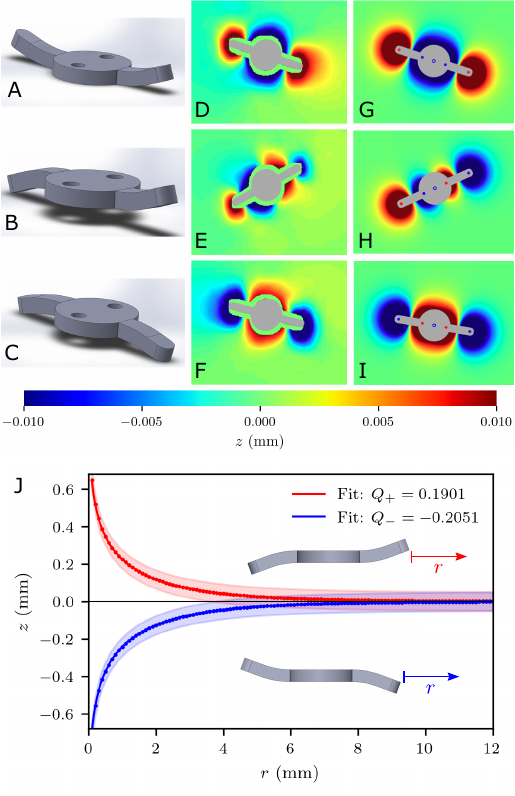}
\vskip -0.2cm
\caption{(A-C) Three physical monomers used to assemble capillary polymers. (D-F) Associated liquid deformation measured by profilometry. (G-I) Simulated interface profiles from the five-charge model of Eq.~(\ref{eq:z_explicit}). (J) Experimental liquid profiles $z(r)$ along the major axis of a $\{^-_+\}$ element and corresponding fits of Eq.~(\ref{eq:z_iso}) near each tip. Shaded bands indicate the uncertainty in the height measurements ($\pm$0.05 mm).}
\label{fig:elements}
\end{figure}

Our system consists of 3D-printed floating monomers with curved branches, whose anisotropy gives rise to more complex interface deformations. Each monomer features a central disk ($D = 8$\,mm diameter, 1.75\,mm thick) with two opposing curved branches ($L = 10$\,mm, $W = 2$\,mm, rounding radius $\varepsilon = 1$\,mm). Each branch $k \in \{1,2\}$ carries a curvature-sign variable $c_k \in \{+1,-1\}$, yielding four logical capillary codes: $\{^+_+\}$, $\{^-_-\}$, $\{^-_+\}$, and $\{^+_-\}$. These reduce experimentally to the three physical monomers shown in Figure~\ref{fig:elements}A-C. 
The interface deformation around each monomer, measured using an improved Schlieren profilometry method \cite{metzmacher2022double,moisy2009synthetic,wildeman2018real}, is shown in Figure~\ref{fig:elements}D-F. The measurements reveal that branches with positive curvature produce upward deformations of the interface near their tips, while negatively curved branches induce downward deformations. Profilometry further shows that the deformation induced by each branch tip is systematically accompanied by an opposite deformation near the branch base, close to the central disk. This compensation reflects the constraint that the total capillary deformation remains governed by the monomer's net weight: while branch curvature redistributes capillary charges locally, any additional deformation must be balanced elsewhere on the monomer.

To model these measured interface deformations, we first recall the simplest isotropic case. For a circular floating object, the interface deformation can be described by a single capillary charge, which characterizes the magnitude and sign of the meniscus curvature and is, in this case, directly linked to the net weight of the object. The liquid elevation $z_i$ at a distance from a floating disk $i$ is given by
\begin{equation}
z_i(\vec{r}) = Q_i K_0\!\left(\frac{|\vec{r} - \vec{r_i}|}{\lambda}\right),
\label{eq:z_iso}
\end{equation}
where $Q_i$ is the capillary charge, representing the characteristic height of the interface deformation \cite{chan1981,kralchevsky2001capillary,vella2005}. A positive charge corresponds to an upward meniscus, while a negative charge results in a downward meniscus. The modified Bessel function of the second kind $K_0$ governs the decay of this deformation over a characteristic capillary length $\lambda = \sqrt{\gamma/\rho_l g}$, which depends on the surface tension $\gamma$, liquid density $\rho_l$, and gravitational acceleration $g$. For the water-air interface, $\lambda \approx 2.7$ mm at room temperature.

To model the anisotropic deformations of our monomers, we employ the Linear Superposition Approximation (LSA), in which the total interface profile is written as a sum of monopolar contributions of the form of Eq.~(\ref{eq:z_iso}) \cite{nicolson1949,kralchevsky2001capillary}. This discrete-charge framework has been used extensively in prior mesoscale capillary self-assembly studies \cite{delens2023,delens2025encoding,delens_3d-printed_2025,vandewalle2020switchable,eatson2024programmable,wang2017dynamic}, and reduces the description of any anisotropic meniscus to a finite set of point charges whose positions and amplitudes are determined by fitting the experimentally measured liquid profiles.
Each branch is thus modeled as a capillary dipole of strength $Q_{b}$, composed of two opposite charges located near the branch tip and near its base. A fifth central charge, $Q_{c}$, accounts for the element's net weight. The total liquid elevation around the monomer $i$ is therefore
\begin{align}
z_i(\vec{r}) =\; &Q_{c}\,K_0\!\left(\tfrac{|\vec{r}-\vec{r}_i|}{\lambda}\right) \nonumber \\
&+ \sum_{k=1}^{2} c_k\,Q_{b}
\Big[K_0\!\left(\tfrac{|\vec{r}-\vec{r}_\mathrm{tip}^{(k)}|}{\lambda}\right)
- K_0\!\left(\tfrac{|\vec{r}-\vec{r}_\mathrm{base}^{(k)}|}{\lambda}\right)\Big],
\label{eq:z_explicit}
\end{align}
with charge positions
\begin{equation}
\vec{r}_\mathrm{tip}^{(k)} = \vec{r}_i + (L-\varepsilon)\,\hat{\boldsymbol{t}}_k, \quad
\vec{r}_\mathrm{base}^{(k)} = \vec{r}_i + \bigl(\tfrac{D}{2}-
\varepsilon\bigr)\,\hat{\boldsymbol{t}}_k,
\end{equation}
where $\hat{\boldsymbol{t}}_k$ points from the centre toward the tip of branch $k$.

Fitting the experimental liquid profiles (Figure~\ref{fig:elements}J) yields $Q_{b+} = 0.1901$ mm and $Q_{b-} = -0.2051$ mm. The slight difference in the magnitudes of these branch charges is attributed to the overall downward deflection caused by the element's net weight, which is captured by the monopolar term $Q_c$. Taking the difference between $|Q_{b+}|$ and $|Q_{b-}|$ gives $Q_{c} = -0.015$ mm. The negative sign reflects the net downward weight of the element, and the small amplitude confirms that branch dipoles dominate the anisotropic deformation. For the following discussions on capillary dimer and polymer, we thus take, as a good approximation, $\lvert Q_{b}\rvert\simeq0.2$ mm for both branch curvatures and $Q_c\simeq-0.075\lvert Q_{b}\rvert$. Using these values, the simulated profiles shown in Figures~\ref{fig:elements}G-I reproduce the sign, location, and relative amplitude of the experimentally observed deformations.

Connecting these monomers produces a polymer whose sequence of branch charges defines the capillary code governing folding.

\subsection*{Pair interactions - Dimer}

When two objects float at the interface, the interaction potential between them, assuming linear superposition holds, is \cite{nicolson1949,kralchevsky1993energetical,vella2005,vassileva2005capillary}
\begin{equation}
U_{ij} = -2\pi\gamma Q_i Q_j K_0\!\left(\frac{|\vec{r}_j - \vec{r}_i|}{\lambda}\right).
\label{eq:pot_2iso}
\end{equation}
For anisotropic elements, this extends to a multipolar formulation summing over all pairs of discrete charges, where $\alpha$ and $\beta$ denote individual point charges belonging to elements $i$ and $j$, respectively:
\begin{equation}
U_{ij} = -2\pi\gamma \sum_{\alpha \in i}\sum_{\beta \in j} Q_\alpha Q_\beta\,K_0\!\left(\frac{|\vec{r}_\beta - \vec{r}_\alpha|}{\lambda}\right).
\label{eq:pot_2aniso}
\end{equation}
Although the LSA neglects contact-line rearrangements and nonlinear meniscus deformation at contact, it provides an effective interaction energy that correctly ranks the observed folded states.
We first analyze the minimal dimer case which consists of $N=2$ monomers linked by a rigid wire and free to rotate. For a typical spacing $d\approx 1$\,mm, set by the connector, the maximum relative angle is
\begin{equation}
\theta_\mathrm{max} = 2\arcsin\!\left(\frac{(D-W+d)/2}{L-\varepsilon}\right) \simeq \frac{\pi}{4}.
\end{equation}

Each branch carries either a positive or negative capillary charge, giving $\mathcal{N} = 4^2 = 16$ possible capillary codes. Despite this diversity, capillary interactions restrict the dimer to only $\Omega = 3$ distinct stable configurations, conveniently described by the discrete state variable $s = \theta/\theta_\mathrm{max} \in \{-1, 0, +1\}$: parallel components ($s = 0$), or tip-contact configurations at $s = \pm 1$.

\begin{figure}
\begin{center}
\includegraphics[width=\linewidth]{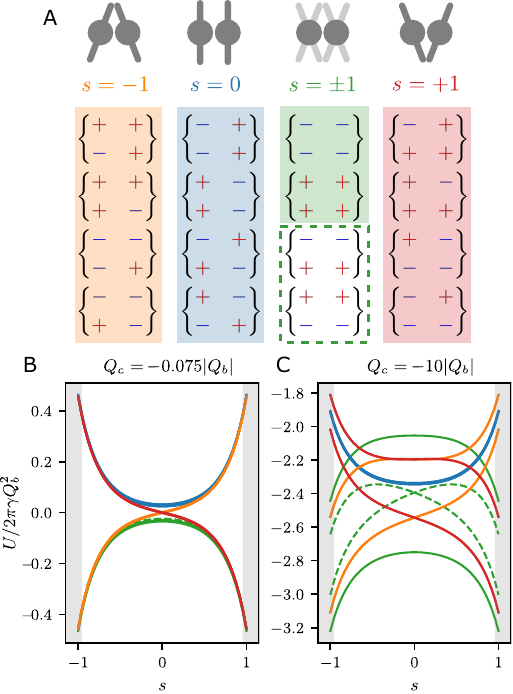}
\caption{(A) The 16 possible capillary codes for a dimer, color-coded by stable state $s \in \{-1,0,+1\}$. Symmetric and asymmetric bistable codes are distinguished (full green and dashed green outlines). (B) Normalized capillary energy landscapes $U(s)$ computed from Eq.~(\ref{eq:pot_2aniso}) using experimental parameters ($Q_{c} = -0.075|Q_{b}|$). (C) Same landscapes computed with an enhanced central charge ($Q_{c} = -10|Q_{b}|$) to highlight symmetry-breaking in the asymmetric bistable cases. Grey areas indicate forbidden regions due to steric contact.}
\label{fig:16codes}
\end{center}
\end{figure}

The 16 capillary codes and their corresponding states are shown in Figure~\ref{fig:16codes}A. Codes with repulsive interactions on both sides lead to $s = 0$; a strong attraction on one side produces $s = \pm 1$; and four codes with attraction on both sides are bistable, allowing either folded state.

An important distinction arises between symmetric and asymmetric bistability. When the capillary code is attractive on both sides but asymmetric with respect to the horizontal axis of the dimer, the two bistable minima are not equally deep: the central negative charge $Q_{c}$ reinforces the deformation and attraction on the negative side, biasing the system toward one state. This symmetry-breaking role of $Q_{c}$ is illustrated in Figure~\ref{fig:16codes}C, where an exaggerated value $Q_{c} = -10|Q_{b}|$ makes the asymmetry explicit.

The energy landscape in Figure~\ref{fig:16codes}B,C also illustrates the important role of the monomer scale. Their millimetric size, comparable to $\lambda$, yields well-defined energy minima. Increasing the object size shifts the effective capillary charges apart and flattens the energy profile, reducing state selectivity. Decreasing the object size, on the other hand, brings opposite charges within distances much smaller than $\lambda$, so that their far-field menisci partially cancel. This capillary-length screening can compromise the reliability of the encoded outcome. The millimetric scale is therefore well suited to exploiting programmable capillary folding.

Having established the interaction rules at the dimer level, we now turn to the collective behavior that emerges when many monomers are assembled into a polymer.

\subsection*{Folding of programmed polymers}
When $N$ monomers are connected by wires and free to rotate, they form a flexible chain. Along its length, branch tips generate alternating capillary charges on both sides, forming two parallel sequences that together constitute the capillary code. 

\begin{figure}[h]
\begin{center}
\includegraphics[width=\linewidth]{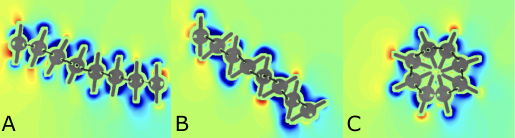}
\vskip -0.2cm
\caption{Experimental liquid profiles around self-folding chains in same color scale as Figure~\ref{fig:elements} (red: positive, blue: negative surface deformations). Three $N = 8$ examples. (A) Linear structure obtained with a straight-biased code. (B) Zigzag when pairs of successive tips attract. (C) Loop formed with a fold-biased code.}
\label{fig:configs}
\end{center}
\end{figure}

Figure~\ref{fig:configs} presents three representative configurations obtained experimentally with $N = 8$. 
A straight line (Figure~\ref{fig:configs}A) emerges when repulsive interactions dominate on both sides, corresponding to $N-1$ consecutive $s = 0$ states, as identified in the dimer analysis.
This configuration is stabilized by persistent repulsion and relaxes back to the linear form under moderate perturbations, as one can observe in Supplementary Movie S1.
A zigzag (Figure~\ref{fig:configs}B) arises when $N-1$ consecutive joints alternate between $s = +1$ and $s = -1$, locking successive branch tips in contact.
A loop (Figure~\ref{fig:configs}C) is programmed by considering repulsions on one side and attractions on the other, yielding $N-1$ consecutive states all equal to $s = +1$ or $s = -1$. 

These three configurations illustrate how a small set of local interaction rules combines to produce qualitatively distinct global geometries. This raises a natural question: as the chain grows, how many such geometries become accessible?
The number of possible capillary codes grows as $\mathcal{N}=4^N$. In contrast, the resulting chain geometry can be described by the state of each joint. Although the joint angle is continuous, the dimer analysis shows that stable configurations occur either in the unfolded, parallel state or at the two folded states. We therefore discretize each joint into three experimentally observed stable states, $s_i=\theta_i/\theta_{\max}\in\{-1,0,+1\}$. Each chain configuration can thus be described by a discrete state vector $\mathbf{s} = (s_1, \ldots, s_{N-1})$. The total number of possible joint microstates would therefore scale as $3^{N-1}$. However, geometric constraints further restrict this space. With a maximal angular deflection $\theta_{\max}\simeq\pi/4$, self-intersection is avoided only up to $N=8$ elements ($2\pi=8\theta_{\max}$). 
For longer chains ($N>8$), self-overlapping configurations must be excluded. 
Furthermore, most of these configurations are not mechanically stable: they correspond to transient states along folding pathways rather than to stable folded structures. Local energy minima are identified numerically by evaluating Eq.~(\ref{eq:pot_2aniso}) over all non-self-overlapping configurations and selecting configurations whose energy is lower than all single-joint variations $s_i \to s'_i$ that are collision-free during the swept rotation. All numerical counts, therefore, refer to the accessible, collision-free discrete transition graph generated by single-joint moves. We distinguish between metastable microstates $\mathcal{M}$, local minima in joint space, and metastable macrostates $\mathcal{G}$ that are geometrically distinct folded structures, free of symmetry degeneracy.

\begin{figure}[ht]
\begin{center}
\includegraphics[width=\linewidth]{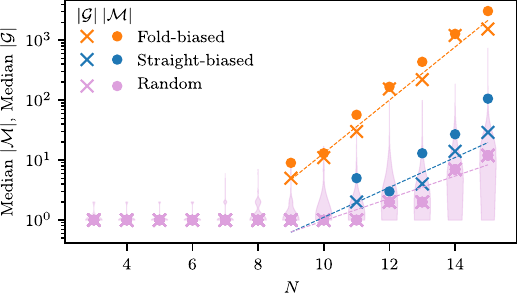}
\caption{Metastable states identified numerically as a function of chain size $N$. Semi-log plot of the median number of metastable microstates $|\mathcal{M}|$ and macrostates $|\mathcal{G}|$ for fold-biased and straight-biased chains, together with the distribution over 1000 random sequences (violin plots). Dashed lines show exponential fits $\sim e^{aN}$ on the median number of metastable macrostates.}
\label{fig:metastable}
\end{center}
\end{figure}

Figure~\ref{fig:metastable} shows the evolution of $|\mathcal{M}|$ and $|\mathcal{G}|$, computed over 1000 randomly generated sequences, as a function of chain length $N$. For short chains ($N < 8$), both quantities remain close to unity: folding is largely deterministic and governed by nearest-neighbor interactions. For these chains, the small variability observed in the violin plot is due to bistable joints as identified in the dimer analysis. For $N \geq 8$, the landscape grows approximately exponentially, with median macrostates scaling as $\sim e^{0.43N}$ for random sequences. Two extreme sequence types bracket this behavior. A fold-biased chain (attractive on one side, repulsive on the other) exhibits the fastest growth ($\sim e^{1.02N}$): because all joints are biased toward bending, successive elements come into close proximity, bringing many capillary charges into competition simultaneously and generating a proliferation of local minima. The fold-biased chain thus sets the upper bound on landscape complexity. A straight-biased chain (repulsive on both sides) grows more slowly ($\sim e^{0.57N}$): although overall curvature is suppressed, partial self-zipping and folding of straight segments can still occur, producing additional minima as $N$ increases. The capillary code, therefore, controls not only the final folded structure but also the complexity of the associated energy landscape. Since $|\mathcal{G}| \ll |\mathcal{N}|$, many different sequences fold to the same geometry: the capillary encoding is inherently degenerate, as we will discuss later.

\begin{figure}
\begin{center}
\includegraphics[width=\linewidth]{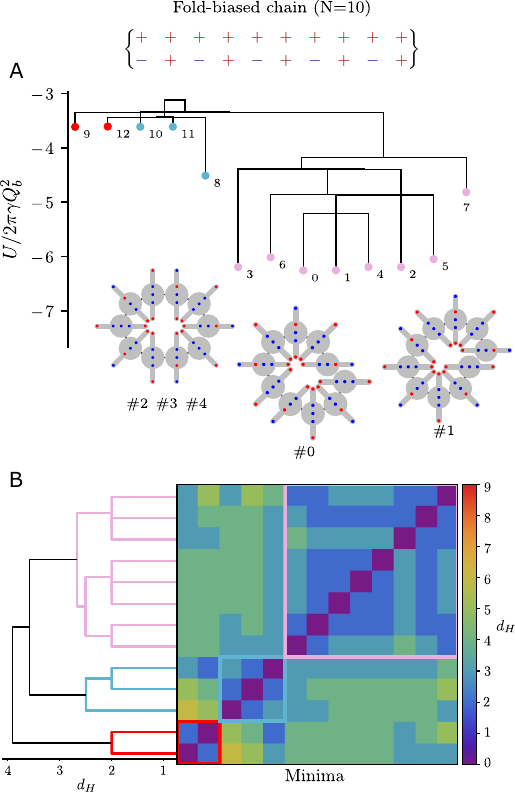}
\caption{Numerical analysis of the landscape of a typical fold-biased chain $N = 10$. (A) Disconnectivity graph with minima ordered by Hamming distance and colored by cluster. The first five lowest-energy microstates are shown; they correspond to only three macrostates. (B) Associated dendrogram and Hamming distance matrix.}
\label{fig:landscape}
\end{center}
\end{figure}

To characterize the structure of these landscapes beyond state counting, we use two complementary tools: disconnectivity graphs, which encode energy barriers between metastable states and reveal the accessibility of folding pathways, and Hamming-based clustering, which groups configurations by structural similarity and identifies mechanically isolated states (singletons). Together, these tools allow the folding landscape to be mapped quantitatively, as illustrated for the fold-biased $N = 10$ chain in Figure~\ref{fig:landscape}A,B. In a disconnectivity graph, each leaf represents a local minimum, and branches merge at the energy level of the lowest barrier connecting the basins of attraction of two minima. In the discrete graph, this barrier is computed as the minimum, over all collision-free single-joint paths connecting two minima, of the maximum energy encountered along the path. The vertical axis represents energy, and the hierarchical structure encodes the accessibility of folding pathways. 
The structural distance between two configurations $\mathbf{s}$ and $\mathbf{s}'$ is quantified by the Hamming distance
\begin{equation}
d_H(\mathbf{s},\mathbf{s}') = \sum_{i=1}^{N-1}\bigl(1 - \delta_{s_i,s'_i}\bigr),
\label{eq:hamming}
\end{equation}
which counts the number of joints whose local state differs. Hierarchical clustering based on $d_H$ groups folded configurations into families; clusters are identified by a threshold $d_H \leq 3$. Microstates that do not belong to any cluster (singletons) correspond to mechanically isolated configurations. The Hamming distance is not a real-space mechanical distance, since changing an early joint can reorganize the entire downstream chain. It nevertheless provides a useful discrete measure of the minimum number of joint-state changes required to separate two configurations. In our dissipative, mechanically driven system, nearby minima in Hamming space are therefore expected to be more easily connected by local rearrangements, whereas distant clusters typically require multiple coordinated joint changes and are mechanically more isolated.

\begin{figure}
\begin{center}
\includegraphics[width=\linewidth]{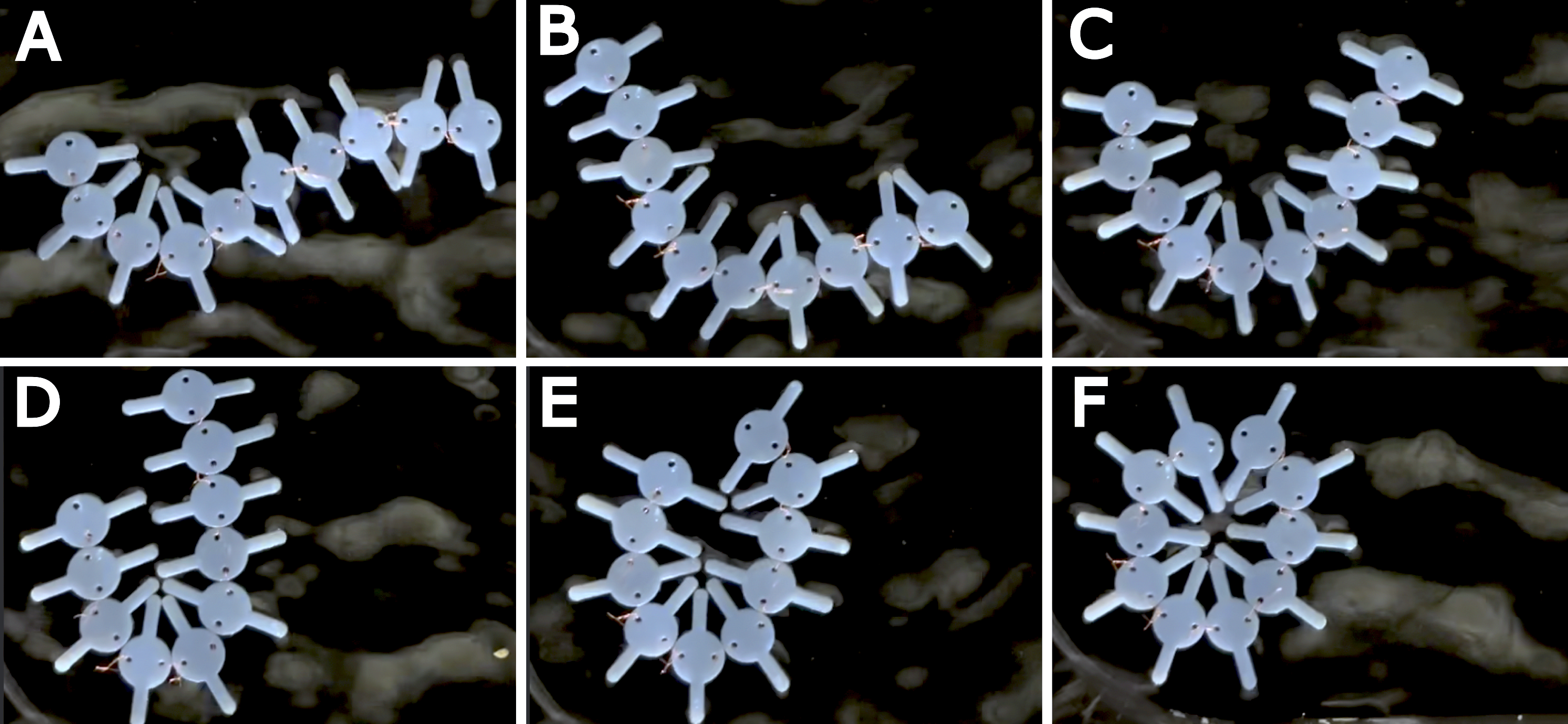}
\caption{Experimental exploration of the folding landscape for a fold-biased chain with $N = 10$ under vertical vibration. (A-F) Selected snapshots from Supplementary Movie S2 showing distinct metastable configurations. Configuration (F) corresponds to a low-energy experimental macrostate, consistent with theoretical predictions.}
\label{fig:N10exp}
\end{center}
\end{figure}
Having mapped the theoretical folding landscape, we now probe it experimentally. In the absence of agitation, capillary bonds between branch tips lock the chain into metastable configurations, and thermal fluctuations alone are insufficient to overcome capillary binding and viscous forces and induce transitions between microstates. However, agitation by Faraday waves and the accompanying turbulent surface flows~\cite{sanli2014,welch2014,xia2019,francois2020} drives structural rearrangements across the folding landscape. We follow the approach of Thomson et al.~\cite{thomson2026nonequilibrium}, who demonstrated that such chaotic surface waves can activate transitions between metastable states in capillary self-assembly, and subject the chain to a vertically vibrated bath (pure water, 6\,mm deep, 13$\times$17\,cm container, vibration frequency $f = 30$ Hz and dimensionless peak acceleration $\Gamma = A(2\pi f)^2/g = 1.2$, where $A$ is the vibration amplitude and $g$ the gravitational acceleration, slightly above the Faraday instability threshold). The chain is observed to transition among multiple metastable configurations, as shown in Figure~\ref{fig:N10exp}, eventually reaching a low-energy macrostate consistent with the lowest-energy minima in Figure~\ref{fig:landscape}.

\subsection*{Sequence robustness and inverse design}
The capillary encoding is inherently degenerate: different capillary sequences can fold into the same equilibrium geometry. This degeneracy is already visible at the dimer level, where multiple capillary codes produce the same final state $s$. For longer chains, $N$-body effects and bistability further accentuate this redundancy, such that a large number of distinct sequences share the same ground macrostate. Such many-to-one mappings between interaction codes and assembled structures are closely related to questions of degeneracy and information capacity in programmable self-assembly~\cite{murugan2015multifarious,huntley2016capacity}.
In this context, a natural question is how robust a given target geometry is to sequence mutations.

An illustrative example is shown in Supplementary Fig.~S1, where replacing a single element of a fold-biased chain can suppress, preserve, or enrich the folding landscape while sometimes leaving the ground macrostate unchanged. To quantify this robustness globally, we define the degeneracy of a target macrostate $\mathcal{G}^*$ as
\begin{equation}
\Xi_N(\mathcal{G}^*) = \bigl|\bigl\{C \in \{1,\ldots,4\}^N : g(C) = \mathcal{G}^*\bigr\}\bigr|,
\label{eq:degeneracy}
\end{equation}
which counts the number of sequences $C$ whose ground state $g(C)$ is identical to $\mathcal{G}^*$. The global prevalence of a target 
structure across sequence space is then $p(\mathcal{G}^*) = \Xi_N(\mathcal{G}^*)/4^N$, estimated here by Monte Carlo sampling over $10^5$ random sequences. To characterize the local mutational robustness around a reference sequence $C_0$, we progressively introduce mutations into the sequence. We define the mutation radius r as the maximum number of elements that are allowed to differ from $C_0$: $r=0$ corresponds to the original sequence, $r=1$ allows any single-element mutation, $r=2$ any combination of two mutations, and so forth. The mutation ball $B_r$ is the set of all sequences within this mutation radius. We then define the mutational robustness $f_{\mathrm{keep}}(r)$ as the fraction of sequences in the mutation ball that still produce the same ground macrostate $\mathcal{G}^*=g(C_0)$.
The size of this mutation ball is $|B_r| = \sum_{k=0}^{r} \binom{N}{k} 3^k$, since each mutated element can take three alternatives. The lower bound for $f_\mathrm{keep}$ is thus $1/|B_r|$, corresponding to the limiting case where only the original sequence $C_0$ itself retains $\mathcal{G}^*$.

\begin{figure}[ht]
\begin{center}
\includegraphics[width=\linewidth]{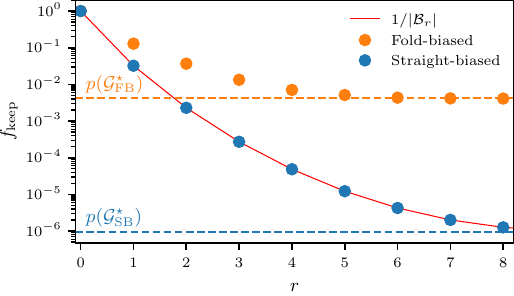}
\caption{Mutational robustness and global degeneracy for $N = 10$. Fraction $f_\mathrm{keep}(r)$ of sequences within mutational distance $\leq r$ from a reference code that produce the same ground-state macrostructure ($\mathcal{G}^*_\mathrm{FB}$ for fold-biased, $\mathcal{G}^*_\mathrm{SB}$ for straight-biased). Dashed lines indicate global probabilities $p(\mathcal{G}^*)$ from Monte Carlo sampling. The red curve is the lower combinatorial limit $1/|B_r|$.}
\label{fig:robustness}
\end{center}
\end{figure}

Figure~\ref{fig:robustness} shows $f_\mathrm{keep}(r)$ for $N = 10$ chains, comparing a fold-biased and a straight-biased reference sequence. The fold-biased chain exhibits a gradual decay: even several simultaneous mutations leave a finite fraction of sequences folding to $\mathcal{G}^*_\mathrm{FB}$, meaning that this fully folded macrostate is not a rare configuration.
An overall curvature of the chain, together with the subsequent collective aggregation of central disks, further reinforces curvature and partially compensates for local perturbations. In contrast, the straight-biased chain follows closely the lower bound $1/|B_r|$, indicating that the straight configuration is highly sensitive to mutations: it is stabilized by local nearest-neighbor repulsion on both sides of each joint, so most mutations destabilize it. The global probability $p(\mathcal{G}^*)$ confirms that the fully folded macrostate is more prevalent across sequence space than the straight configuration. The gap between the two curves in Figure~\ref{fig:robustness} thus quantifies the contrast in programmability between robust and fragile target structures.

This redundancy of the capillary encoding mirrors biological coding, where distinct genetic sequences can give rise to similar structural motifs. It also has direct implications for design: highly degenerate target structures are intrinsically robust, whereas unique targets require precise sequence control.

\section*{Conclusions}
We have demonstrated flexible capillary polymers whose folding is encoded by the geometry of their monomers. For short chains, folding is largely deterministic and governed by nearest-neighbor interactions, yielding straight, zigzag, or loop conformations. Around $N\simeq8$, the
onset of geometrical self-contact enables increasingly nonlocal interactions, and for longer chains the folding landscape becomes increasingly complex, with the number of competing metastable states growing approximately exponentially with chain length. We further show that the capillary sequence determines not only the folded geometry, but also the degeneracy and mutational robustness of the resulting structures. Together, these results establish capillary polymers as a quantitative mesoscale model of sequence-dependent folding and a controllable platform for studying the statistical physics of folding landscapes.

More broadly, these results place capillary polymers within recent efforts to realize sequence-programmed folding in colloidal systems~\cite{mcmullen2022programmable}. Here, geometry provides a fixed interaction alphabet, so that increasing chain length naturally introduces nonlocal couplings when distant segments of the backbone come into spatial proximity. Simple local folding rules can thereby give rise to a rapidly expanding set of competing metastable states. Capillary polymers thus provide a complementary setting for exploring how competing interactions and geometrical constraints shape programmable assembly landscapes~\cite{zeravcic2014size}.

Several perspectives follow from this work. Dynamic reconfiguration between folded states could be achieved by actuating individual branches using magnetic \cite{surfaceswimm2018,sunlocally}, magnetoelastic \cite{poty2017magnetoelastic}, shape-shifting \cite{cappello2023bioinspired}, or elastocapillary \cite{annurev:/content/journals/10.1146/annurev-fluid-122316-050130} properties. In particular, a time-varying sequence of folded states propagating along the chain would generate a traveling body wave at the liquid interface, reminiscent of the metachronal undulation used by polychaete worms for propulsion~\cite{daniels2021tomopteris,burns2024polychaete}. Such capillary undulatory locomotion at the millimeter scale has recently been demonstrated with magnetically actuated flexible sheets~\cite{ren2024undulatory}, suggesting that programmable folding chains could serve as a segmented, sequence-controlled analog. Finally, the bistable joints identified in the dimer analysis offer an unexploited degree of freedom for storing binary information along a chain. In the present system, the small ratio $|Q_{c}/Q_{b}|=0.075$ limits the symmetry breaking between the two bistable minima. Designing elements with a larger central charge would accentuate the asymmetric bistability identified in Figure~\ref{fig:16codes}C, selectively biasing each bistable joint toward one of its two states. Beyond programmable self-assembly, capillary polymers provide a versatile macroscopic platform for investigating how sequence-encoded local interactions give rise to complex folding landscapes, structural robustness, and emergent functionality.
\section*{Materials and Methods}
\subsection*{Design and 3D printing}
The elements have been designed with SolidWorks and printed using a PolyJet 3D printer (Stratasys J35 Pro). The PolyJet method consists of jetting out resin droplets, which are then cured using UV light. It results in an announced resolution of up to 16 $\mu$m. We used a resin similar to ABS plastic (Vero Blue) in a glossy finish. The printing process induces a surface-finish asymmetry: the lower face, which is in contact with the support material during printing, has a matte finish, whereas the surfaces that do not contact the support material remain glossy. In all experiments, the monomers were placed with the matte face toward the water and the glossy face toward the air. The contact line was observed to pin near the perimeter separating these two surface finishes, likely because of their difference in wettability. Chains were assembled by connecting neighboring elements with a metallic wire of diameter $0.25$ mm and then twisting it to secure the connections. The connection fixed the center-to-center distance, while allowing relative rotation between adjacent elements. The typical spacing between neighboring disks was $d\simeq1$ mm. 
\subsection*{Experimental procedure}
All experiments were performed at room temperature using pure water. To deposit the chains at the air-water interface while minimizing disturbances of the free surface, the assembled chain was first placed on a rigid mesh positioned above the water surface. The mesh was then lowered very slowly into the bath. As the mesh passed below the interface, the chain remained trapped at the air-water surface by capillary forces. This procedure allowed the chains to be deposited with minimal perturbation of the interface and of their initial configuration.
\section*{Data availability}
STL files for the monomers are openly available in the Zenodo repository at \href{10.5281/zenodo.18697355}{https://zenodo.org/records/18697356}. All study data are included in the article and/or supporting information.
\section*{Acknowledgements}

This work was financially supported by the FNRS CDR project J.0186.23 ``Magnetocapillary Interactions for Locomotion at Liquid Interfaces'' (MILLI). A.F. is a Research Fellow of the Fonds de la Recherche Scientifique-FNRS. N.V. thanks the Fondation Francqui for support.

\bibliography{biblio}
\newpage
\onecolumngrid

\begin{center}

{\Large\bfseries Supporting Information}

\vspace{0.8cm}

{\LARGE\bfseries Capillary self-folding chains}

\vspace{0.8cm}

{\large
Megan Delens, Axel Franckart, Martin Poty, and Nicolas Vandewalle
}

\vspace{0.4cm}

GRASP, Physics Department, Universit\'e de Li\`ege, Belgium.

\vspace{0.4cm}

Corresponding author: \texttt{megan.delens@uliege.be}

\end{center}

\vspace{1cm}

\clearpage


\section{Sensitivity to single-element mutations}
\label{sec:supp_mutations}

\begin{figure}[ht]
    \centering
    \includegraphics[width=0.7\linewidth]{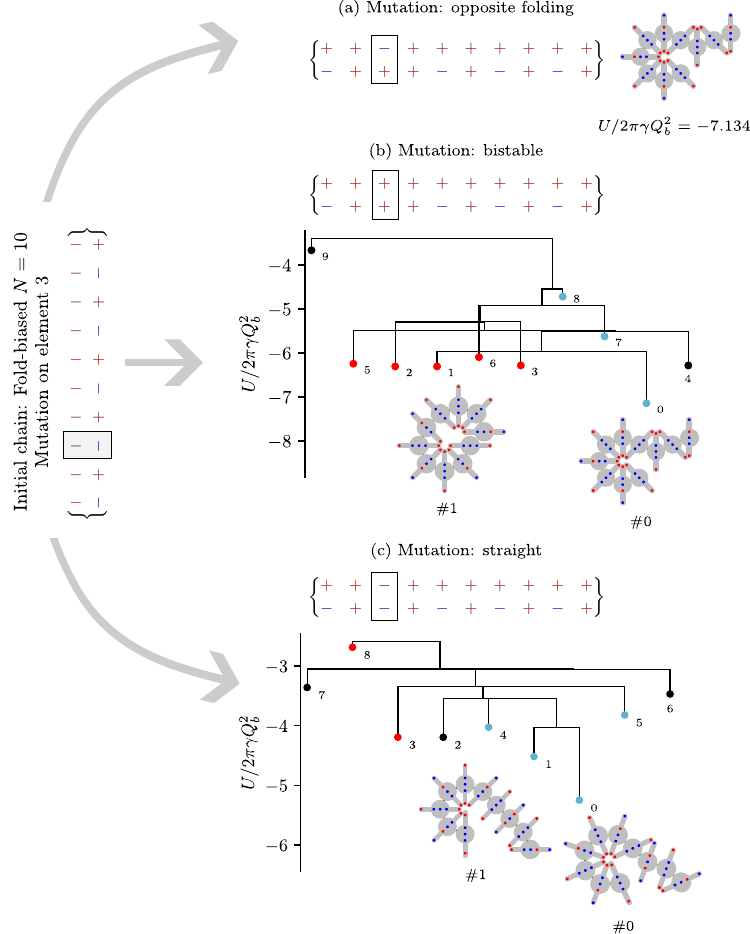}
    \caption{
    Mutation effect on a fold-biased chain $N=10$.
    Disconnectivity graphs and first minimal configuration(s)
    for mutation on element 3.
    (a) Opposite-folding mutation: one metastable microstate.
    (b) Bistable mutation: nine microstates in small Hamming clusters.
    (c) Straight mutation: overall chain curvature still induces
    folding in the ground state.
    }
    \label{fig:supp_mutations}
\end{figure}

Figure~\ref{fig:supp_mutations} shows the effect of replacing
a single element (position 3) in a fold-biased $N=10$ chain.
Three representative mutations are considered:
(a) a mutation inducing opposite folding tendency in adjacent
joints which suppresses alternative pathways, leaving a single
metastable microstate;
(b) a bistable mutation which introduces nine microstates
organized into small Hamming clusters;
(c) a repulsion-inducing mutation which reduces local curvature
but still leads to a folded ground state driven by the overall
chain curvature and central-disk attraction.
It is worth noting that both the bistable and opposite-folding
mutations lead to the same ground-state configuration,
illustrating the degeneracy of the capillary encoding.

\clearpage


\section{Supplementary videos}
\label{sec:supp_videos}

\textbf{Movie S1.}
Straight ($N=8$) chain: spontaneous relaxation to the linear
configuration after perturbation, and response under
Faraday-wave agitation.

\vspace{0.5cm}

\textbf{Movie S2.}
Fold-biased chain ($N=10$) under vertical vibration:
transitions between metastable configurations and eventual
trapping near the ground macrostate.
\end{document}